**ACN WHITE PAPER**

# Agentic Communication Networks: When Agents Become Network Entities

**Authored by**

**Renwei (Richard) Li**
Southeast University, China

**Biwei Li**
Southeast University, China

**Nadjib Aitsaadi**
Paris-Saclay University, France

**Tarik Taleb**
Ruhr University Bochum, Germany

**Cedric Westphal**
University of California, Santa Cruz, USA

**Tao Huang**
Purple Mountain Laboratories, China

**Luca Foschini**
University of Bologna, Italy

**Zuqing Zhu**
University of Science and Technology of China, China

# CONTENTS

# Agentic Communication Networks: When Agents Become Network Entities

## ABSTRACT

Autonomous agents are moving from application-layer components to active participants in networked systems. They not only generate traffic but also perceive context, express intent, invoke services, coordinate with other agents, and act under task, policy, and resource constraints. This evolution exposes a mismatch between emerging agentic interactions and conventional endpoint- and traffic-centric network abstractions.

This white paper introduces Agentic Communication Networks (ACN) as a network-centric paradigm for the time when agents become first-order network entities. At its core, ACN reshapes networking around three fundamental aspects. First, agents are represented as identifiable and governable network entities with explicit identity, authority, and lifecycle. Second, agentic services are expressed through intent, capability, context, policy, and task state, allowing the network to capture what an agent seeks to accomplish and under what conditions. Third, agentic execution becomes an explicit network concern as tasks involve multiple agents, services, tools, and distributed capabilities across heterogeneous infrastructures. Based on these shifts, the paper defines the scope, principles, architecture, and research directions of ACN, positioning it as a networking foundation for autonomous agent ecosystems.

# 1. Introduction: When Agents Become Network Entities

Communication networks have continuously evolved as the entities they connect have changed. Early networks primarily connected terminals, hosts and networks. Mobile networks extended connectivity to users and personal devices. The Internet of Things further expanded the network edge to sensors, machines, vehicles, cameras, meters, robots, and embedded systems. More recently, cloud, edge computing, digital twins, and cyber-physical systems have made services, data, and computing resources integral parts of networked operation. Each transition did not merely increase the number of connected endpoints; it changed the assumptions under which networks were named, addressed, managed, optimized, and secured [1]-[7].

Autonomous agents represent the next such transition. An agent is not simply another application that sends and receives traffic. It can observe context, maintain state, reason over objectives, invoke tools, interact with services, negotiate with other agents, and initiate actions on behalf of a human user, an enterprise process, a machine, a robot, or an infrastructure system. An AI-powered travel agent may coordinate calendars, payment services, airline systems, and user preferences. An industrial inspection agent may combine sensing data, quality-control policies, digital-twin states, and robotic actions. A network-operation agent may diagnose faults, request measurements, reconfigure resources, and coordinate with other management agents.

This change creates a new networking problem. Existing networks are highly effective at connecting endpoints, forwarding packets, supporting sessions, isolating slices, enforcing access policies, and providing quality-of-service treatment. However, these mechanisms were not designed with the agent as an explicit network entity. A host can be addressed, a user can be authenticated, a device can be attached, an application flow can be classified, and a service can be exposed. Yet an agent may not be identical to any one of these objects. It may run across multiple devices, migrate between edge and cloud, act for different principals under different authorities, invoke multiple tools, delegate subtasks to other agents, and participate in long-lived task execution across organizational or administrative boundaries.

As a result, agentic communication cannot be fully characterized by conventional descriptions of traffic alone. The source and destination of a packet remain important, but they do not answer several network-relevant questions raised by agentic operation: Which agent is acting? On behalf of whom? Under what authority? For which task? With what intent? Which capability is requested? Which policy constraints apply? Which resources are involved? Which actions should be recorded, verified, or audited? These questions are not external to networking once agents begin to coordinate tasks through distributed infrastructures. They affect service discovery, access control, routing, resource allocation, privacy protection, reliability, safety, accountability, and trust.

The importance of these questions can be seen from simple scenarios. If a personal agent asks a cloud service to retrieve private records, the network may need to distinguish whether the request is made by the authorized user, by a delegated agent, or by an untrusted process imitating that agent. If two industrial agents coordinate robotic actions, low latency alone is insufficient unless the agents are authorized, the task context is valid, and the action path satisfies safety policies. If an edge agent requests computation for a perception task, the relevant service requirement is not merely bandwidth or delay, but whether the task can be completed with sufficient confidence, within a deadline, and under privacy constraints. These cases show that agentic communication introduces service semantics that are closely coupled with network operation.

This observation does not imply that the network should freely inspect application payloads or infer private user intentions. On the contrary, a rigorous agentic network architecture must preserve the separation between private content and network-visible service expressions. What should become visible is not an agent's complete reasoning process or arbitrary payload meaning, but explicitly exposed, authorized, and policy-controlled information needed for service support. Such information may include agent identity, delegated authority, task class, service intent, capability requirement, context descriptor, policy constraint, trust attribute, or task state. The key design question is therefore not whether the network should capture all aspects of an agent, but which agent-related information should be represented, exposed, validated, and consumed by authorized network functions.

This is the point at which Agentic Communication Networks become necessary. ACN is motivated by the recognition that future networks will need native abstractions for autonomous agents, just as previous network generations developed abstractions for hosts, users, devices, flows, sessions, services, and slices. ACN does not aim to replace the Internet, mobile networks, cloud platforms, or agent frameworks. Nor is it limited to using artificial intelligence to manage networks. Instead, ACN asks a more fundamental question: how should communication networks evolve when autonomous agents become persistent, identifiable, and governable participants in networked task execution [8]-[12].

These shifts establish a bidirectional relationship: networks expose connectivity, computing, sensing, data, and service capabilities to agents, while agents contribute reasoning and adaptation to networked systems. ACN goes further: agents become network entities, tasks become service-level objects, and heterogeneous capabilities become composable resources. The network entity must therefore expand beyond hosts, devices, users, and applications; the service expression beyond addresses, flows, and QoS parameters to controlled intent, context, capability, policy, and task state; and the service objective beyond connectivity and delivery toward trusted task coordination.

These shifts define the central position of this white paper. When agents become network entities, communication networks must move beyond treating agent traffic as opaque application traffic. They must provide principled mechanisms for identifying agents, representing authorized intent, discovering capabilities, coordinating resources, enforcing policies, and maintaining accountability across heterogeneous infrastructures. ACN aims to support large-scale agent ecosystems by enabling connectivity, interoperability, trust, task coordination, and network-safe operation, including protection against uncontrolled task propagation and excessive resource consumption.

The remainder of this white paper develops this position. Section 2 defines the scope of ACN and clarifies its boundaries with respect to agent frameworks, semantic communication, AI-assisted network management, and existing network architectures. Section 3 presents the foundational shifts introduced by agentic communication. Section 4 introduces a reference architecture for ACN. Section 5 discusses potential network capabilities designed to support agents as network entities. Subsequent sections examine service models, candidate enabling mechanisms, deployment scenarios, open research challenges, and standardization opportunities. Together, these discussions aim to establish ACN as a network-centric foundation for the emerging era in which autonomous agents become first-order participants in communication systems.

## 2. Definition and Scope of ACN

The introduction has motivated ACN based upon a network centric observation: autonomous agents are no longer merely application components that generate traffic over existing infrastructures; they are emerging as network-relevant entities that participate in distributed task execution. This section defines ACN, clarifies what constitutes an agent in the context of ACN, identifies the scope of agentic communication, and distinguishes ACN from adjacent research areas. The purpose is to establish a precise conceptual boundary before discussing architectures, capabilities, service models, and standardization opportunities.

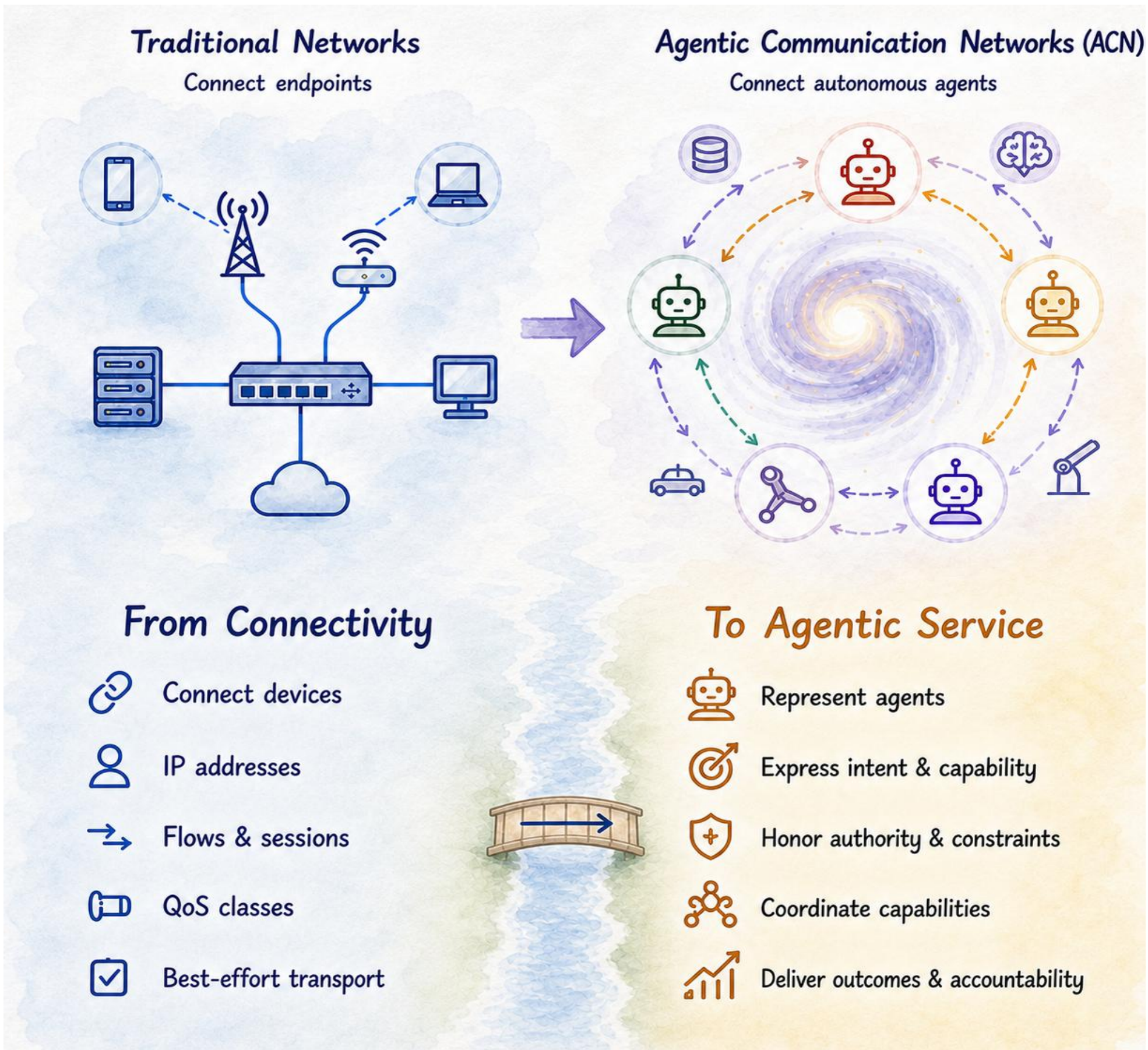

Fig. 1. From Connectivity-Centric Networks to Agentic Communication Networks.

### 2.1. Definition of Agentic Communication Networks

**Agentic Communication Networks (ACN)** refer to communication networks that provide native support for autonomous agents as first-order network entities and enable their trusted, intent-aware, policy-compliant, and task-oriented interactions across heterogeneous communication, computing, sensing, storage, data, and service infrastructures.

More specifically, ACN aims to support autonomous agents that can be identified, addressed, authenticated, authorized, discovered, coordinated, and governed within networked environments. Unlike conventional networking models that primarily focus on connecting endpoints and delivering traffic, ACN introduces agent-related abstractions into the network service model. These abstractions may include agent identity, delegating authority, service intent, task context, capability requirements, execution constraints, policy conditions, trust attributes, and task states. Through these abstractions, ACN enables the network to provide controlled support for agent discovery, intent-aware service interaction, capability matching, resource coordination, policy enforcement, and accountable task execution.

This definition emphasizes four essential properties.

First, ACN is **network-centric**. Its primary concern is not how an individual agent reasons, plans, learns, or invokes tools internally, but how agents are represented, connected, coordinated, and governed when they interact through communication networks.

Second, ACN is both **agent-aware** and **network-aware**. It treats agents as entities with network-relevant identity, authority, capability, state, and accountability, rather than reducing them to opaque application processes or traffic sources.

Third, ACN is **intent-aware** but policy-bounded. It allows agents to expose service intent, task context, and constraints in standardized and controlled forms, but does not assume unrestricted network access to private application content or arbitrary semantic interpretation of payloads.

Fourth, ACN is **task-oriented**. Its service objective is not limited to packet delivery or QoS provisioning. Instead, ACN supports the coordination of communication, computing, sensing, storage, and service resources required for agentic task execution under trust, policy, and resource constraints.

## 2.2. Agents as Network Entities

In ACN, an agent is defined from the perspective of communication networking, with emphasis on its network-visible identity, authority, interactions, capabilities, and task participation.

ACN encompasses both digital agents and physical agents. Digital agents are software-based autonomous entities operating in devices, edge platforms, cloud systems, applications, or network functions. Examples include personal assistants, service agents, network management agents, digital-twin agents, and enterprise agents. Physical agents are autonomous or semi-autonomous entities whose decisions and interactions are associated with physical sensing, mobility, control, or actuation, such as robots, vehicles, drones, industrial machines, and other embodied systems. Despite their different forms, both types of agents may participate in distributed tasks, invoke networked capabilities, interact with other entities, and act under explicit authority and operational constraints.

Accordingly, an agent in ACN is defined as an autonomous or semi-autonomous network-relevant entity that can act on behalf of a human, organization, machine, service, application, or infrastructure and participate in networked task execution through interactions with other agents, services, tools, devices, or network functions.

Agents may exhibit different levels of intelligence, autonomy, embodiment, and persistence. Their implementation may involve large language models, control algorithms, planning systems, sensing and perception modules, or combinations of these components. Their relevance to ACN arises from the network-level requirements created by their identity, authority, capabilities, interactions, and task dependencies.

An agent becomes a network entity when its operation creates network-visible relationships that must be represented or managed. These relationships may include the association between an agent and its principal, the authority under which it acts, the capabilities it requests or provides, the intent and task context associated with its interactions, and its dependencies on distributed resources or other entities. An agent may therefore need to be identified, authenticated, authorized, discovered, coordinated, monitored, or audited throughout task execution. Under these conditions, an IP address, device identity, or service endpoint alone cannot fully represent the agent and its role in the network.

## 2.3. Agentic Communication

Agentic communication is defined as network-supported interactions in which autonomous agents exchange or expose information required for task execution, service invocation, resource coordination, policy compliance, or collaborative decision-making. It is broader than packet transmission, message exchange, or API invocation because it also captures the task context, intent, authority, state, and coordination semantics associated with an interaction, as well as how such information is interpreted and acted upon across agents and network functions.

In conventional networking, communication is often described in terms of source and destination addresses, traffic types, transport sessions, application protocols, QoS parameters, and service-level requirements. In ACN, these descriptions remain necessary but may not be sufficient. Agentic communication may additionally involve several network-relevant expressions. Here, a network expression refers to information about an agent, task, or interaction that can be represented and exposed to network functions to support communication, coordination, control, or policy enforcement.

The first is identity expression, which specifies which agent is acting and under which principal, credential, role, or authority. The second is intent expression, which describes a controlled, network-relevant representation of the task objective or service goal. The third is capability expression, which specifies what function, tool, model, sensor, actuator, data source, or service is requested or provided. The fourth is context expression, which captures task-relevant conditions such as location, time, environment, network state, service state, or operational mode. The fifth is policy expression, which specifies constraints related to privacy, security, safety, compliance, trust, cost, or administrative domain. The sixth is task-state and result expression, which represents progress, dependency, intermediate or final results, confidence, failure conditions, or coordination state during task execution.

## 2.4. Agentic Networking: Network Support for Agents

Agents as network entities define who participates in networked operation, while agentic communication defines the task-relevant information exchanged among agents, services, tools, and network functions. Agentic networking addresses the corresponding network question: how should the network represent, expose, and coordinate its capabilities so that these interactions can be supported under explicit service, policy, trust, and resource constraints? Agent reasoning may motivate a request, but network support is realized through network-facing abstractions and enforceable mechanisms.

Such support extends beyond carrying agent-generated traffic. The network may need to authenticate the acting agent and its authority, accept a controlled service expression, discover capabilities without requiring a predetermined destination, map task constraints to communication and infrastructure resources, and preserve task-relevant relationships as paths, nodes, providers, or execution locations change. These functions may be implemented through APIs, signaling, control-plane state, service gateways, programmable data planes, or combinations of existing and emerging mechanisms across heterogeneous infrastructures.

While new protocols are needed for ACN, some protocols exist that could be leveraged and adapted. For instance, BPP (Big Packet Protocol) is a possible starting point for a network-layer realization within this design space [13]-[15]. ACN may require an extensible mechanism for representing addressing and delivery information, service requirements, metadata, processing instructions, and payload-related information in a structured and independently extensible form. BPP is one candidate approach that can support such requirements. An ACN-aware BPP profile could therefore bind a packet or packet group to selected network-relevant expressions, such as an Agent ID, Task ID, requested capability or service class, execution constraint, information role, validity interval, or bounded task state. This would allow authorized network functions to apply task-aware forwarding, scheduling, processing, or assurance without reconstructing the agent's complete reasoning process.

The qualitative-communication model developed with BPP further shows how network treatment can reflect the role of information within a task. Packet Wash allows a forwarding node to remove less significant payload chunks under constrained conditions while preserving timely delivery of more important information [15]. Similar treatment may be useful when an agentic task exchanges observations, features, model state, inference results, and control actions with different values and deadlines. BPP is nevertheless only a candidate mechanism: agent-related information need not appear in every packet, and any data-plane use requires minimum disclosure, integrity, authorization, bounded semantics, and graceful fallback.

Recent work in qualitative and semantic-aware communication provides a concrete basis for this form of network support. Qualitative communication can assign significance to payload chunks and, under constrained conditions, discard lower-value chunks instead of an entire packet so that the receiver still obtains timely, useful information [39]. Semantic-aware multiple access extends the principle across users: the concepts of self-throughput and assisted-throughput represent how one successful transmission can satisfy both the transmitting user and other users with correlated semantic segments, enabling distributed access decisions that balance utilization and fairness [40]. For ACN, these results suggest that a task-aware data-plane expression may need to represent information significance, bounded substitutability, redundancy relationships, and the metric used to assess task usefulness. Such metadata must remain authorized, integrity-protected, and limited to what network functions require.

## 2.5. Scope of ACN

The scope of ACN includes the network principles, architectures, protocols, interfaces, service models, and governance mechanisms required to support agents as first-order network entities. It covers both control-plane and data-plane functions and spans heterogeneous communication, computing, sensing, data, and service infrastructures. ACN also interfaces with application-layer agent systems, edge and cloud platforms, programmable networks, digital twins, and cyber-physical systems.

As shown in Fig. 2, the scope of ACN can be characterized through three fundamental architectural gaps introduced by agentic operation: the entity gap, the service gap, and the execution gap.

**At the entity level**, ACN addresses how autonomous agents become identifiable, governable, and persistent participants in networked systems. This includes agent naming and identification, authentication and authorization, agent–principal relationships, authority and delegation, registration and discovery, lifecycle management, mobility, and accountability. Related mechanisms already exist in specific domains, such as industrial networks, where devices such as PLCs can maintain persistent identities and operate with predefined roles and permissions. ACN extends these ideas toward more general, autonomous, and heterogeneous agents that may move across devices, sessions, platforms, and administrative domains while retaining verifiable identity, authority, and accountability.

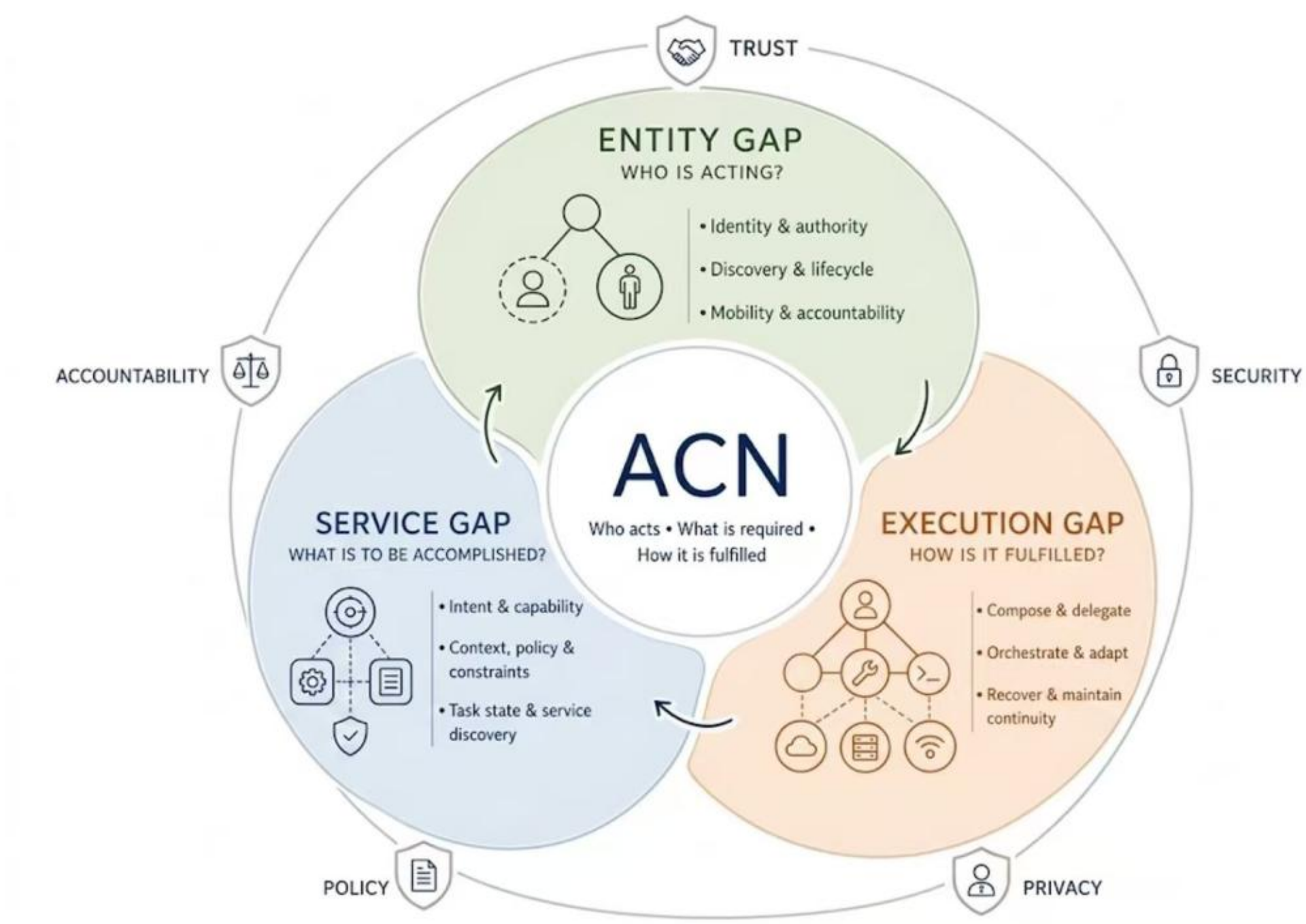

Fig. 2. Three Architectural Gaps Motivating Agentic Communication Networks.

**At the service level**, ACN addresses how the objectives and requirements of agents are represented in forms that can be interpreted and supported by the network. Relevant abstractions include intent, capability, context, policy, constraints, and task state. They enable agents to request or provide capabilities, discover appropriate services, establish task-related interactions, and express service requirements beyond conventional endpoint and traffic descriptors. This scope includes agent-to-agent, agent-to-service, agent-to-tool, agent-to-machine, agent-to-human, and agent-to-network interactions.

**At the execution level**, ACN addresses how agentic tasks are carried out across multiple agents, services, tools, network functions, and distributed capabilities. Task execution may involve different participants, execution environments, administrative domains, and resource types over time. ACN therefore studies capability composition, task decomposition and delegation, service binding, orchestration, multi-agent interaction, communication–computing–sensing integration, execution-state management, adaptation, failure recovery, and task continuity. The execution scope connects agentic service requirements with the networked capabilities that collectively support their fulfillment.

Trust, security, privacy, policy, and accountability span all three levels. Agent identity must be trustworthy, service expressions must be authorized and selectively exposed, and task execution must remain policy-compliant and auditable across participating entities and domains. These requirements create cross-cutting research problems in trust establishment, privacy-preserving information exposure, cross-domain authorization, provenance, safety enforcement, and accountable execution.

Thus, ACN defines a networking framework for representing who is acting, what is to be accomplished, and how agentic tasks are executed across distributed networked capabilities. Its scope provides the foundation for the reference architecture and native network capabilities developed in the following sections.

# 3. Reference Architecture of ACN

The ACN reference architecture describes how networks can support autonomous agents without prescribing a protocol stack, platform, access technology, or deployment model. It provides a common view through which heterogeneous agents interact with heterogeneous networked resources while preserving boundaries among agent intelligence, network coordination, and infrastructure execution [16]-[23].

ACN addresses these requirements through three logical planes: the Agent Entity Plane, the Agentic Service Plane, and the Network Coordination Plane. These planes are built above existing communication and computing infrastructures. They do not replace routing, transport, resource management, cloud platforms, edge systems, sensing control, or application services. Their purpose is to provide the agent- and task-related abstractions through which these existing capabilities can participate coherently in agentic operation.

The three planes answer three fundamental architectural questions:

- **Who is acting?**
- **What service relationship is being established?**
- **How can the network support it?**

The Agent Entity Plane establishes the actor and its authority. The Agentic Service Plane establishes the requested service relationship. The Network Execution Plane establishes the executable arrangement through which that relationship can be fulfilled. Together, they extend network architecture from the management of endpoints and traffic toward the support of identifiable agents, capability-oriented services, and task-level coordination.

## 3.1. Three Logical Planes of ACN

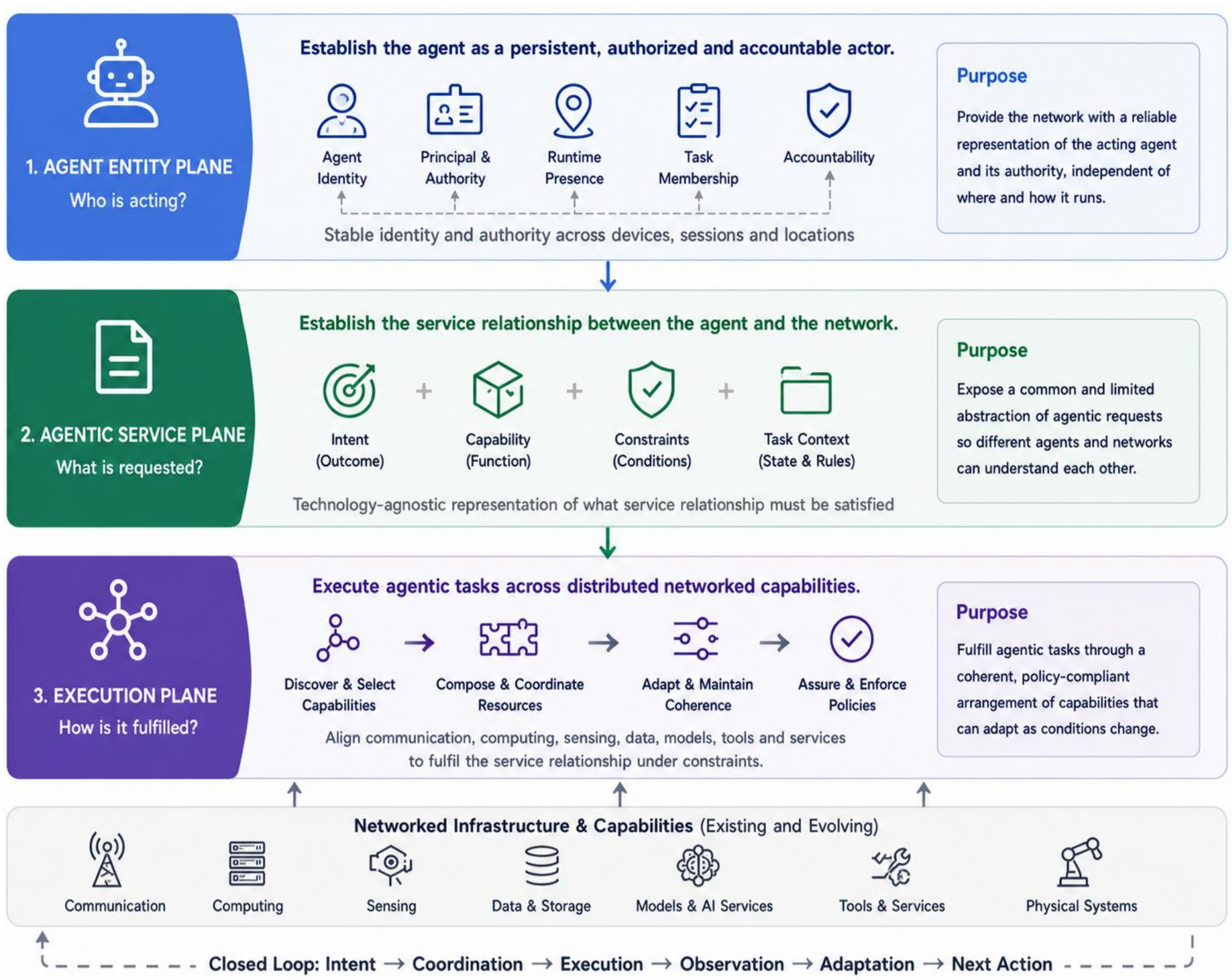

Fig. 3. Three-Plane ACN Architecture for Agent Identity, Service Representation, and Network Execution.

As illustrated in Fig. 3, we define a three-plane ACN architecture consisting of the Agent Entity Plane, the Agentic Service Plane, and the Agentic Execution Plane. The Agent Entity Plane establishes the actor and its authority. The Agentic Service Plane establishes the requested service relationship. The Agentic Execution Plane establishes how that relationship is carried out across distributed agents, services, tools, and networked capabilities. Together, these three planes extend network architecture from the management of endpoints and traffic toward the support of identifiable agents, capability-oriented services, and distributed agentic execution.

**Agent Entity Plane：** The Agent Entity Plane provides a network-visible representation of the agent. It identifies which agent is acting, which principal the agent represents, what authority has been granted to it, and how long that authority remains valid. This representation allows the network to distinguish the agent from the device, application, or connection currently carrying its traffic.

Such separation is important because an agent may move between a device, an edge node, and a cloud platform, while multiple agents may also share the same physical device. A physical agent such as a robot may retain the same agent identity while changing network attachments, and a digital agent may continue the same task after its execution environment changes. The Agent Entity Plane therefore maintains the identity, authority, lifecycle, and accountability of the acting agent across these changes.

**Agentic Service Plane：** The Agentic Service Plane describes what an agent wants to accomplish in a form that can be supported by the network. It represents the agent's intent together with the capabilities it requests or provides, the constraints that must be satisfied, and the task context in which the interaction takes place. These elements form a common service expression between the agent and the network.

For example, an inspection agent may request a sensing capability within a specified area, before a deadline, and under a privacy constraint. The agent does not need to specify in advance which sensor, edge server, communication path, or service provider will be used. The Agentic Service Plane captures the required outcome and conditions, while exposing only the information needed for network support.

**Network Execution Plane：** The Agentic Execution Plane determines how the requested service is actually carried out. A task may involve one agent using several distributed capabilities, or multiple agents, services, tools, models, sensors, and network functions participating at different stages. The execution plane identifies suitable participants and capabilities and organizes their roles, dependencies, and service bindings during task execution.

This process may involve capability discovery, task decomposition, delegation, orchestration, routing, computing placement, sensing activation, QoS control, and resource allocation. The execution arrangement can also change as agents join or leave, providers become unavailable, network conditions vary, or task requirements evolve. The task identity, intent, and constraints can remain valid while the underlying agents, paths, services, and resources are reconfigured, allowing ACN to maintain task continuity across a distributed and dynamic execution environment.

# 4. Native Network Capabilities

## 4.1. Agent Registration and Discovery Capabilities

ACN provides unified agent registration, identity resolution, capability discovery, and status synchronization. It assigns each agent a unique identifier and maintains an up-to-date directory of agent capabilities, service interfaces, network locations, and operational states. By dynamically handling agent registration, de-registration, revocation, and status changes, including the removal or suspension of compromised or non-compliant agents, the network can maintain a unified namespace and controlled entry point for agent interconnection [9], [24].

### 4.2. Service Orchestration Capability

ACN automatically organizes and coordinates multiple agents according to service requirements. By considering agent capabilities, resource availability, service priorities, and network conditions, it dynamically generates task execution plans and supports the on-demand composition and coordinated scheduling of agent services, thereby improving multi-agent collaboration in complex scenarios [16], [17], [20], [21].

### 4.3. QoS Assurance Capability

ACN dynamically allocates communication resources according to application requirements for latency, bandwidth, reliability, and jitter. By leveraging an optical-electrical converged deterministic network, it establishes end-to-end deterministic communication channels for real-time applications such as robotic control, coordinated drone operations, and vehicle-road collaboration, ensuring reliable transmission and service continuity for agent collaboration [25].

### 4.4. Protocol Translation Capability

ACN aims to facilitate interoperability among heterogeneous agents by providing common mechanisms for capability description, information exchange, and semantic translation, while recognizing that complete interoperability may not always be achievable because of capability mismatches or ambiguous and non-equivalent representations. It provides unified mappings across communication protocols, data formats, and capability interfaces while supporting protocol translation, message encapsulation, data adaptation, and capability proxying. By abstracting underlying protocol and platform differences, ACN provides unified interfaces for upper-layer agent collaboration [26]-[29].

### 4.5. Trusted Security Capability

ACN provides lifecycle-wide security and accountability for agent interconnection through identity authentication, access control, secure communication, trusted computing, resource accounting, operational auditing, and verification. It can record and attribute agents' use of network, computing, storage, and service resources, while enabling selected actions and outputs to be checked against policies, provenance information, task requirements, or other verification mechanisms when needed. These capabilities protect agent access, capability invocation, data transmission, and collaborative execution, while supporting trusted identities, communications, behaviors, data, and accountable resource use in large-scale open agent ecosystems [30]-[33].

## 5. Enabling Mechanisms and Candidate Technologies

### 5.1. Automatic Agent Registration and Discovery

Automatic registration and discovery provide a unified and trustworthy view of agent resources. Each agent is assigned an Agent ID associated with its identity, organization, network location, credentials, capabilities, service interfaces, resource requirements, and operational status. Heartbeat reporting and event-driven updates maintain dynamic Agent Profiles.

Standardized capability ontologies, knowledge graphs, and semantic embeddings are used to normalize heterogeneous capability descriptions. Vector retrieval, Graph-RAG, and LLM-based intent analysis map task requirements to suitable agents. Candidate agents are ranked according to capability relevance, real-time load, service quality, reliability, reputation, and prior collaboration performance.

Candidate technologies include decentralized identifiers (DIDs), digital certificates, capability ontologies, knowledge graphs, vector databases, RAG, Graph-RAG, graph neural networks, and LLM-based semantic matching.

## 5.2. Multi-Agent Collaborative Orchestration

Collaborative orchestration transforms service requirements into executable multi-agent workflows. An LLM-assisted hierarchical planner decomposes a complex objective into tasks, subtasks, and atomic operations and represents their dependencies, constraints, resource demands, and evaluation criteria as a Task Graph.

Agents are selected and assigned roles such as planner, coordinator, executor, and supervisor according to their capabilities, QoS attributes, trust levels, resource availability, and collaboration history. During execution, the system continuously monitors task progress, communication quality, and resource consumption. It can reschedule tasks, replace failed agents, reconfigure the collaboration topology, and optimize resource allocation in response to runtime changes.

Candidate technologies [34-37] include LLM-based task planning, task knowledge graphs, graph-based workflows, constraint optimization, multi-agent reinforcement learning, graph neural networks, workflow engines, and dynamic scheduling algorithms.

## 5.3. Cross-Domain Scheduling and QoS Assurance

Cross-domain QoS assurance coordinates agent capabilities, computing resources, network conditions, and service costs over an optical-electrical converged deterministic network. Agent gateways maintain an Agent Forwarding Table that associates each semantic service identifier with candidate agent instances, reachable paths, network conditions, computing loads, and service prices [38].

LLM-based intent understanding converts natural-language requirements into routable semantic service identifiers. After filtering candidates against mandatory constraints, the gateway jointly evaluates latency, bandwidth, jitter, reliability, computing capacity, geographic location, and cost to select the most appropriate agent instance and network path. Hierarchical filtering and Top-N pruning reduce decision overhead for large candidate sets and enable millisecond-level scheduling.

Candidate technologies may include IPv6 anycast identifiers, SRv6, deterministic networking, active network probing, telemetry, semantic addressing, vector and relational retrieval, multi-objective optimization, weighted scoring, and candidate-pruning algorithms.

Semantic-aware orchestration also requires lifecycle management for the knowledge used to interpret network-visible expressions. The KB-MANO framework treats the refinement, placement, updating, and distribution of shared knowledge bases as joint communication-computing decisions and demonstrates, through a proof of concept, that knowledge-management overhead and radio-resource allocation are coupled [41]. In ACN, analogous shared state may include capability ontologies, semantic embeddings, protocol mappings, task models, or policy vocabularies. These artifacts require explicit version and compatibility identifiers, freshness and locality constraints, distribution policy, and resource accounting so that an agentic service can be rebound or migrated without silently changing the meaning of its intent, capabilities, or task state.

## 5.4. Heterogeneous Protocol Translation

Heterogeneous protocol translation enables semantic interoperability among MCP, A2A, ACP, and domain-specific agent protocols. An Agent Exchange Gateway translates each source protocol into a Protocol-Independent

Intermediate Representation (PIIR) that uniformly represents message metadata, task semantics, capabilities, context, interaction state, and security policies.

The translation process consists of protocol parsing, PIIR generation, target-protocol reconstruction, and runtime validation. Protocol knowledge graphs, semantic mappings, and template engines preserve task intent and interaction state during conversion. A plug-in architecture allows a new protocol to be integrated by defining only its bidirectional mapping to the PIIR, reducing the complexity of supporting $N$ protocols from $O(N^2)$ pairwise mappings to $O(N)$.

Candidate technologies include abstract syntax trees, protocol state machines, semantic ontologies, protocol knowledge graphs, embedding-based alignment, LLM-assisted parsing, schema mapping, template engines, protocol plug-ins, and runtime consistency verification.

Work on Model Context Protocol integration for the Internet of Robotic Things provides one implementation example of semantic decoupling at the edge. In the reported prototype, heterogeneous nodes expose sensor data as MCP resources and actions as tools, allowing an MCP host and a language model to assemble workflows without a bespoke adapter for every service [42]. The same evaluation also identifies a boundary that is important for ACN: semantic interpretation and tool invocation account for much of the workflow latency, so MCP complements rather than replaces deterministic middleware for strict real-time or safety-critical control. An ACN translation gateway should therefore preserve typed capability schemas, authority, task context, and latency class across protocols, while keeping time-critical execution on a bounded and verifiable control path.

### 5.5. Trusted Security Assurance

Trusted security establishes lifecycle-wide protection covering agent identity, capabilities, behavior, data, communication, and results. A zero-trust mechanism (ZTM) combines unique identities, certificates, cryptographic signatures, and dynamic trust evaluation to provide trusted access and reputation-based authorization. The ZTM includes policy engines, runtime monitoring, risk-aware controls, and standardized audit.

Policy engines enforce least-privilege access to tools, data, and other agents. Runtime monitoring detects prompt injection, unauthorized access, malicious tool invocation, abnormal collaboration, and sensitive-data leakage. Risk-aware controls can dynamically restrict privileges, isolate an agent, suspend execution, or terminate collaboration. Standardized audit logs and tamper-resistant records preserve task execution, delegation, data access, and security-event provenance for accountability.

Candidate technologies include DIDs, Public Key Infrastructure (PKI), digital signatures, multifactor authentication, zero-trust access control, capability-based authorization, policy engines, input/output guardrails, anomaly detection, security knowledge graphs, trusted computing, confidential communication, blockchain-based evidence preservation, and trusted timestamps.

## 6. Use Cases and Deployment Scenarios

### 6.1. Dynamic Collaboration

**Scenario:** A city is preparing a temporary, accessible public event. A civic coordinator agent assembles a transit shuttle, an accessibility assistant, a venue service robot, a food-logistics drone, and a local energy controller. The participants belong to different owners and platforms. They join only for the mission, assume bounded roles, and leave as their subtasks finish.

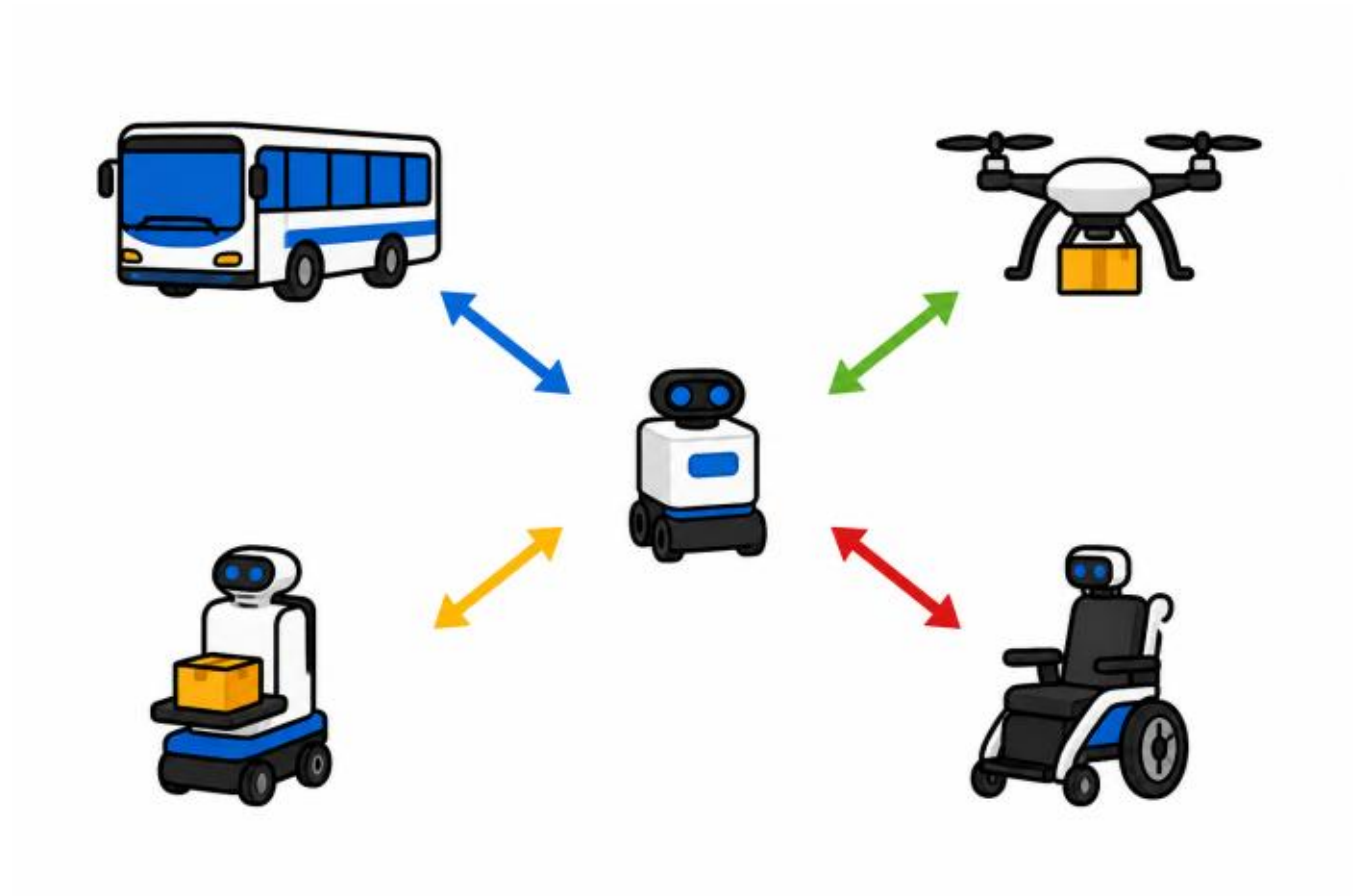

Fig. 4. Scenario view of dynamic collaboration.

**Why ACN is Needed:** Conventional connectivity can carry messages between endpoints, but it does not establish whether an autonomous participant is entitled to join, which capability it is offering, or how its actions relate to the shared outcome. ACN makes the temporary team a network-visible service relationship: identities and authority are verified, capabilities are matched to subtasks, communications follow changing team membership, and completion evidence remains attributable.

**Representative operational flow**

| 01 | 02 | 03 | 04 | 05 |
|---|---|---|---|---|
| Declare the mission, constraints, location, deadline, and accessibility policy. | Discover candidate agents and verify identity, ownership, authority, and capability claims. | Bind selected agents to roles; create a task-scoped communication group and policy context. | Coordinate dependencies and timing; admit or replace participants as the plan evolves. | Verify subtask outcomes, release resources, and close the temporary trust relationship. |

**Key ACN requirements**

- Persistent agent identity independent of device or session
- Capability registration, discovery, and freshness
- Task-scoped authorization and least-privilege membership
- Dynamic group communication and role-aware policy
- Accountable state, handover, and completion evidence

## 6.2. Intent-Driven Services

**Scenario:** An environmental field agent is asked to inspect a remote coastline before an approaching storm and return a trustworthy assessment. The requester states the desired outcome and constraints—not a fixed network configuration. The task may require a rover, coastal sensors, edge inference, intermittent satellite access, and a human escalation path, with the mix changing as weather and coverage evolve.

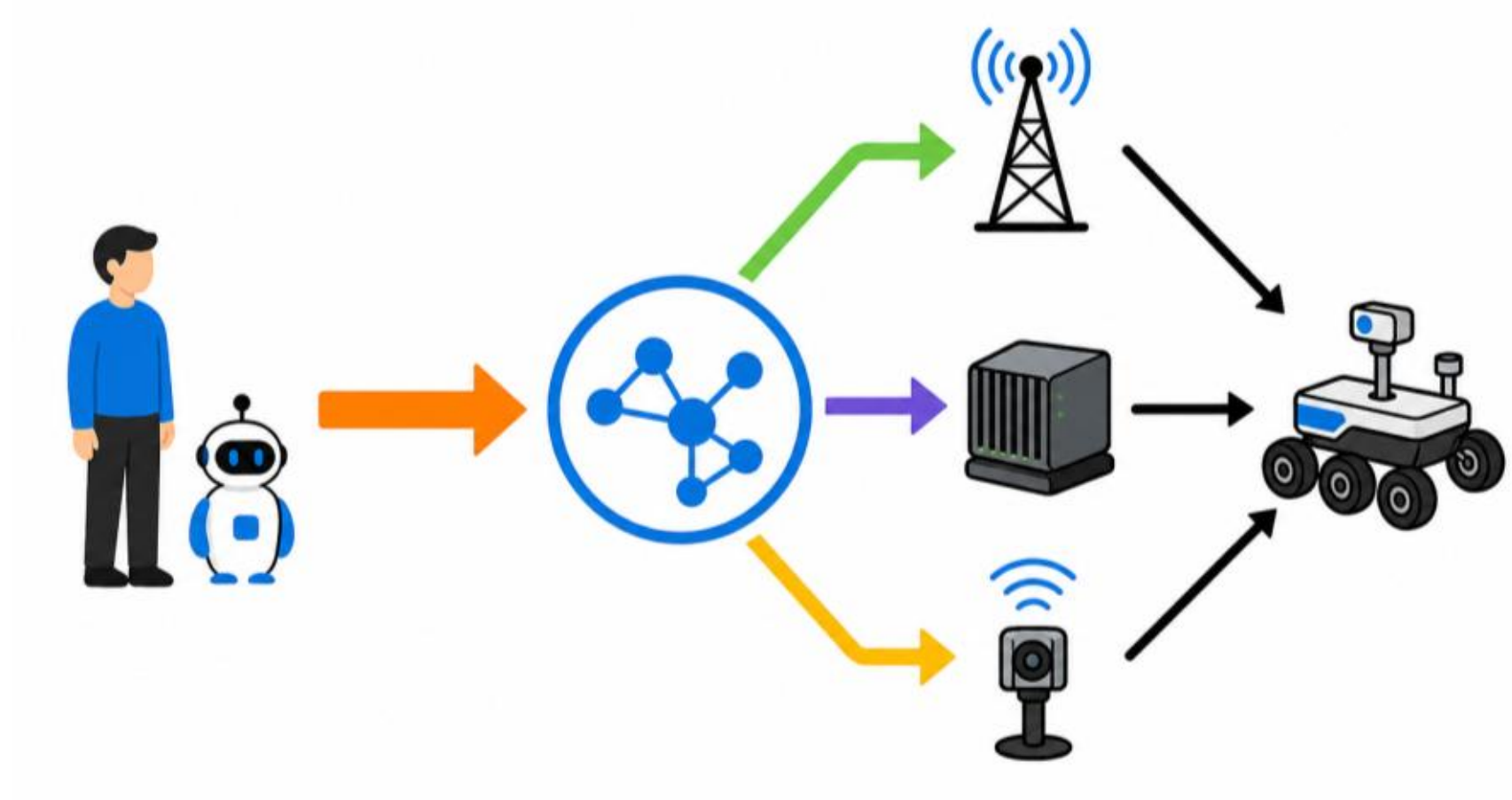

Fig. 5. Scenario view of intent driven services.

**Why ACN is Needed:** No single service invocation captures this outcome. ACN interprets intent, tests feasibility, discovers suitable capabilities, and maintains the relationship between the requested result and its changing realization. When radio conditions deteriorate or a sensor becomes unavailable, the system can negotiate a degraded-but-useful alternative instead of merely failing an isolated session.

**Representative operational flow**

| 01 | 02 | 03 | 04 | 05 | 06 |
|---|---|---|---|---|---|
| Express the intended outcome, evidence quality, geographic scope, deadline, and policy constraints. | Translate intent into capability requirements and dependency options. | Check feasibility; select providers and negotiate an acceptable service composition. | Activate communications, sensing, computing, and agent services as one task context. | Observe progress and conditions; adapt the composition or propose an alternative. | Verify the outcome against the original intent and preserve provenance. |

**Key ACN requirements**

- Structured intent, constraints, and outcome semantics
- Capability abstraction across communication, compute, sensing, data, and agents
- Feasibility assessment and alternative negotiation
- Closed-loop observation, adaptation, and fulfilment verification
- End-to-end provenance across a changing service composition

### 6.3. Cross-Domain Cooperation

**Scenario:** A regional outage isolates part of a community. An operator network, a municipal emergency network, and a utility/logistics network each retain partial connectivity, sensing, and transport capacity. Their agents form a temporary resilience federation so responders, connected citizens, drones, and repair crews can obtain essential service across administrative boundaries.

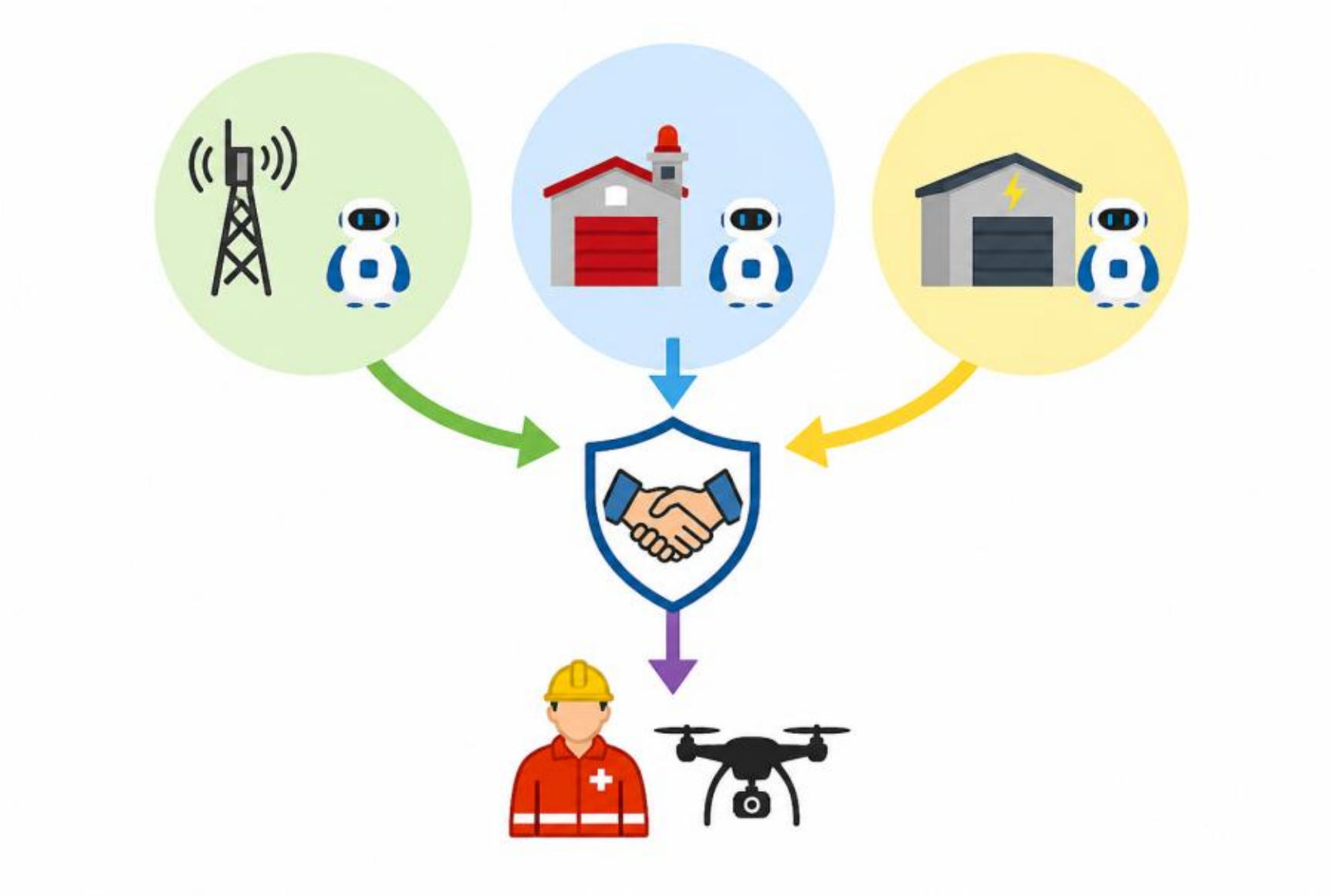

Fig. 6. Scenario view of cross-domain cooperation.

**Why ACN is Needed:** Static agreements cannot anticipate every combination of failure, load, geography, and policy. ACN provides interoperable agent identity, explicit delegated authority, machine-readable service offers, and policy-constrained negotiation. The federation can expose only the minimum required capabilities, preserve each domain's control, and continuously revise the arrangement as capacity returns.

**Representative operational flow**

| | | | | | |
|---|---|---|---|---|---|
| 01<br>Detect a continuity gap and publish a bounded request for assistance. | 02<br>Discover trusted peer agents and exchange verifiable credentials and policy metadata. | 03<br>Negotiate area, duration, priority, capacity, accounting, and permissible data use. | 04<br>Create a temporary federated service and steer eligible task traffic through it. | 05<br>Observe load and mission impact; rebalance resources without breaking task continuity. | 06<br>Repatriate services and terminate temporary authority when recovery permits. |

**Key ACN requirements**

- Federated identity, credentials, and trust anchors
- Policy and authority negotiation across administrative domains
- Interoperable capability and service descriptions
- Task-priority continuity with privacy and data-minimization controls
- Auditable decisions, accounting, revocation, and graceful exit

### 6.4. Task-Aware Data Exchange

**Scenario:** A drone, ground robot, and edge agent inspect a wind farm together. Early in the task, the drone sends dense imagery for scene understanding. The ground robot later exchanges compact features for localization. Near a suspected fault, a small inference and control update become more urgent than the bulk imagery that produced them.

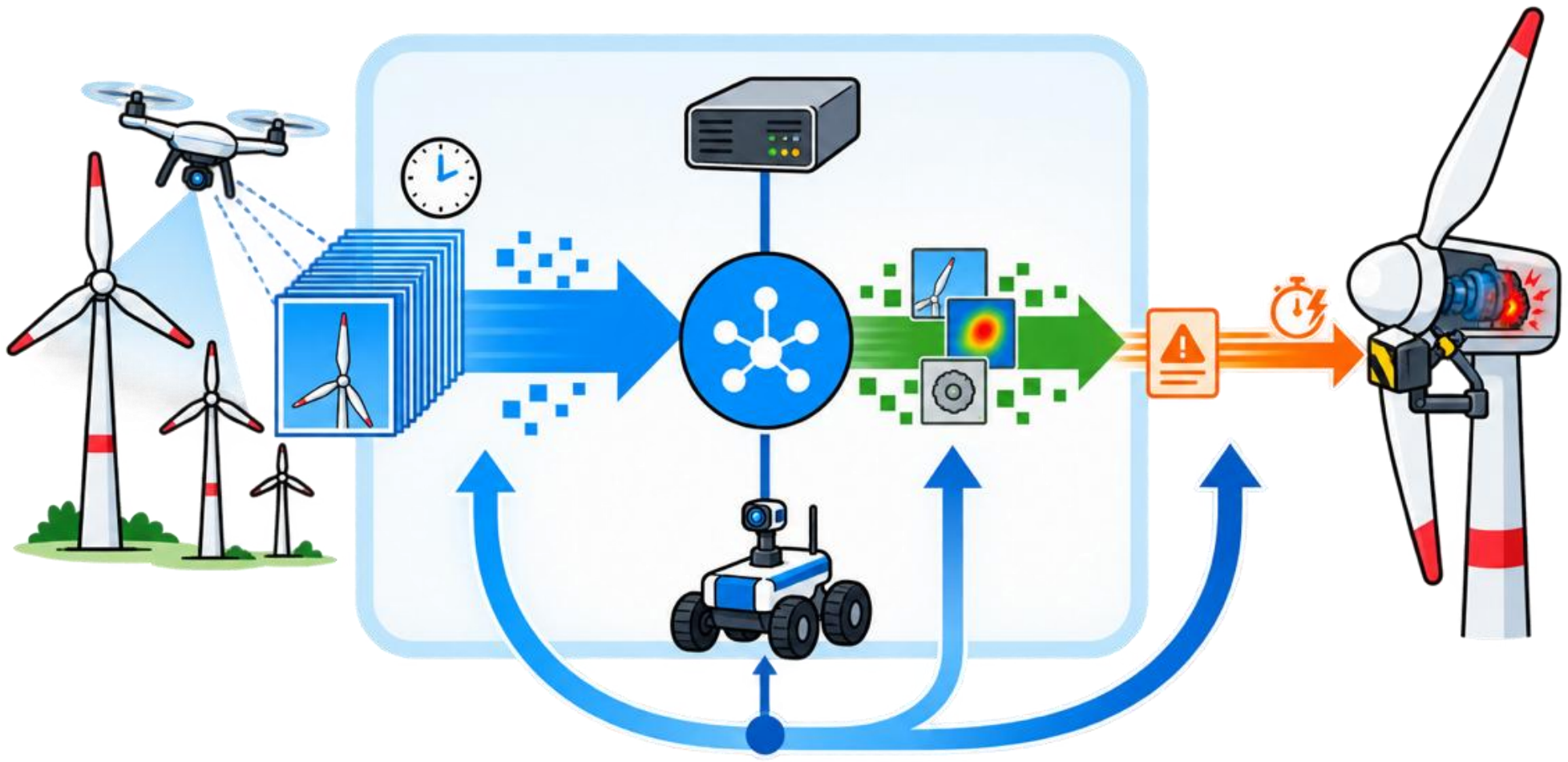

Fig. 7. Scenario view of task-aware data exchange.

**Why ACN is Needed:** Packets from the same agents can have very different value, deadlines, loss tolerance, and privacy sensitivity depending on task stage. Endpoint and application labels alone cannot express that evolution. ACN associates exchanges with a task, information role, validity window, and dependency—allowing the network to prioritize actionable results, reduce redundant raw transfer, or relocate processing while preserving task correctness.

Recent studies illustrate how task value and inter-agent redundancy can alter radio-resource decisions. In a multi-UAV coverage setting, semantic-aware power allocation optimizes the quality of the server's reconstructed observation rather than raw throughput, and the learned policy depends on the selected perception metric [43]. In vehicular uplink, a two-stage design uses V2V exchange to estimate overlap among camera fields of view before V2I transmission prioritizes semantically important, non-redundant patches; its results also expose a tradeoff between time spent discovering redundancy and time available for uplink delivery [44]. PRISM generalizes redundancy-aware access to variable-length, asynchronous transmissions with fairness and energy-efficiency objectives, using self-throughput and assisted-throughput so that devices can act from local observations [45].

For ACN, this evidence supports a task-aware exchange contract that can identify the information unit, its task role, the applicable reconstruction or utility metric, known redundancy relationships, confidence and validity limits, and the energy or cost budget. Suppression, substitution, or partial delivery should be logged and reversible through a fallback to ordinary delivery when overlap estimates, calibration, encryption, model compatibility, or policy prevents reliable reuse. The reported gains are based on simulation or prototype settings, including idealized reconstruction assumptions in parts of the vehicular evaluation, so these mechanisms should be treated as candidate ACN capabilities whose task-level benefit must be verified in deployment [43]-[45].

**Representative operational flow**

| | | | | | |
|---|---|---|---|---|---|
| 01 Associate data producers and consumers with a shared inspection task. | 02 Advertise information form, freshness, confidence, dependency, and handling constraints. | 03 Choose where to transform raw observations into features or inference results. | 04 Apply stage-aware bandwidth, latency, reliability, multicast, and protection treatment. | 05 Update priorities when the task enters diagnosis or control; suppress stale or redundant data. | 06 Record which evidence informed the final maintenance action. |

**Key ACN requirements**

- Task and subtask association for information exchanges
- Semantic role and lifecycle: observation, feature, model state, inference, command
- Deadline, freshness, confidence, dependency, and privacy metadata
- Adaptive processing placement and group-aware delivery
- Traceability from data products to decisions and actions

## 6.5. Industrial Embodied Agents

**Scenario:** A flexible microfactory produces customized components in short runs. A mobile manipulator, vision-quality agent, digital-twin/edge agent, safety sentinel, and human supervisor collaborate. When a robot becomes unavailable or a quality anomaly appears, the cell replans work, moves inference, and reroutes material without violating safety or traceability constraints.

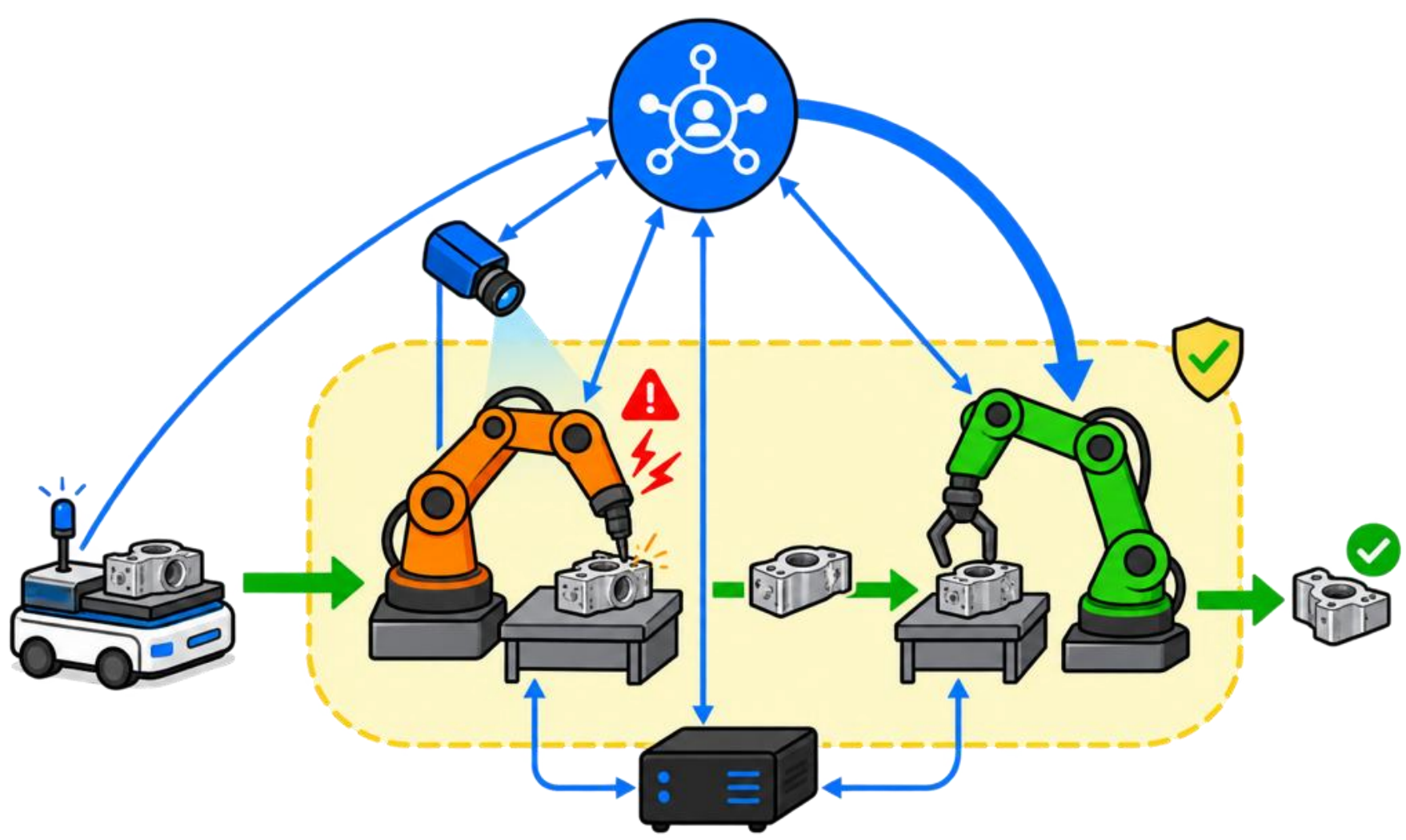

Fig. 8. Scenario view of industrial embodied agents.

**Why ACN is Needed:** Embodied agents convert network decisions into physical action. Coordination therefore cannot optimize communications in isolation: sensing freshness, model state, control deadlines, workspace permissions, and human oversight must remain coherent. ACN provides a task-level control context that can admit agents, bind responsibilities, enforce safety constraints, and adapt the end-to-end arrangement while the production objective continues.

**Representative operational flow**

| 01 | 02 | 03 | 04 | 05 | 06 |
|---|---|---|---|---|---|
| Register agents, tools, workspaces, safety envelopes, and current operational state. | Translate the production order into an executable task graph with dependencies and checkpoints. | Bind roles and reserve communication, edge compute, sensing, and physical resources. | Execute with synchronized perception, control, verification, and human oversight. | Detect degradation; isolate unsafe components and recompose roles, paths, or processing placement. | Verify quality and provenance; close authority and archive accountable task evidence. |

**Key ACN requirements**

- Embodied-agent identity, attestation, and lifecycle state
- Real-time capability and workspace descriptions
- Safety-aware authorization and human override
- Deterministic coordination across communication, compute, sensing, and actuation
- Fast failure isolation, substitution, and state-consistent recovery

# 7. Research Challenges and Opportunities

The transition from conventional communication networks to agentic communication networks (ACN) requires more than introducing AI models into network operation or exposing additional network APIs to autonomous applications. Once agents become network entities, the network must support new relationships among agents, principals, authorities, intents, capabilities, tasks, resources, and outcomes. The fundamental challenge is to establish these relationships in forms that are technically meaningful to the network, interoperable across heterogeneous systems, and sufficiently stable for long-term deployment.

The reference architecture developed in this white paper identifies three corresponding architectural planes: the Agent Entity Plane, the Agentic Service Plane, and the Network Execution Plane. Each plane exposes a distinct set of open problems. The Agent Entity Plane requires a network representation of autonomous actors that can persist across devices, sessions, and administrative domains. The Agentic Service Plane requires a common abstraction through which agents can express network-relevant intent and capabilities without exposing unrestricted application semantics. The Network Execution Plane requires mechanisms that can translate these service relationships into coherent use of distributed communication, computing, sensing, data, and service capabilities.

## 7.1. Agent Entity: From Device Identity to Agent Identity

Current communication systems possess mature mechanisms for identifying subscribers, devices, applications, network functions, and service endpoints. An autonomous agent introduces a different type of entity. Its logical existence may persist while its execution moves among a device, an edge platform, and a cloud environment. Multiple agents may share one device, while one agent may simultaneously use multiple devices, sessions, and network connections. An agent may also act on behalf of another agent, a human, an enterprise, or an infrastructure operator.

The first research challenge is therefore to determine what constitutes a network-visible agent identity. Binding an agent permanently to an IP address, device identifier, application instance, or model instance would make its identity dependent on transient execution conditions. At the other extreme, assigning long-lived identities indiscriminately to short-lived agent instances could introduce unnecessary complexity as well as privacy and management concerns. ACN therefore needs an identity model that distinguishes the logical agent from its current runtime presence and supports different identity lifetimes, scopes, and execution environments.

A second challenge concerns the relationship among identity, authority, and accounting. Knowing which agent initiated an interaction does not establish whether that agent is permitted to perform the requested action, nor how the resources consumed by that action should be attributed. An agent may act on behalf of a principal under delegated authority that is limited by capability, task, time, location, resource, or administrative domain. Delegation may also be chained when one agent assigns a subtask to another. In such cases, resource consumption may span multiple network locations, devices, services, and sessions, requiring usage records to be associated with the relevant agent, principal, and delegated task and reconciled across these contexts. ACN therefore requires mechanisms for representing the identity of the acting agent, the authority under which an action is performed, and the corresponding resource usage and accountability.

This issue becomes particularly important when agents interact across administrative domains. Two agents may belong to different users, enterprises, operators, or service platforms and may have no prior relationship. The network must establish sufficient trust for interaction without assuming a universal agent platform or a single identity provider.

Research is consequently needed on persistent and scoped agent identities, agent–principal binding, delegated authority, multi-agent delegation chains, privacy-preserving identification, agent mobility across execution environments, revocation, and accountability. A central question is how much of this information must be visible to the network and how much can remain within application and security systems.

The corresponding standardization opportunity lies in defining interoperable network-facing representations of agents while leaving their internal implementation unconstrained. Candidate areas include scoped agent identifiers and namespaces, agent–principal associations, authority and delegation descriptors, agent registration and discovery procedures, lifecycle states, authentication and authorization interfaces, and mechanisms for associating agent identity with existing device, user, and service identities.

The long-term objective is a simple property:

**An agent should remain a recognizable and accountable network entity even when its endpoint, runtime, or communication session changes.**

## 7.2. Agentic Service: From Traffic Requirements to Task-Oriented Service

Conventional networks are designed around well-defined service descriptors such as addresses, traffic classes, bandwidth, latency, reliability, and application identifiers. Agents commonly operate at another level of abstraction: they pursue outcomes, request capabilities, impose contextual constraints, and revise their actions according to intermediate results.

The challenge is not simply to make the network "understand intent." Such a formulation is too broad to provide an architectural foundation. A more productive way of phrasing the question is:

**What is the minimum network-relevant information required to support an agentic task?**

An agent may possess extensive private context, reasoning history, memory, goals, and application state. Most of this information should never be required by the network. ACN therefore needs to identify a narrow and stable boundary between application semantics and network-relevant service semantics.

The service abstraction developed in this white paper begins with five concepts: agent, intent, capability, constraint, and task context. Considerable research remains necessary to determine their precise semantics and relationships.

Intent must be expressive enough to capture the desired service outcome while remaining interpretable by heterogeneous networks. Capability descriptions must allow function-oriented discovery without becoming application-specific catalogs. Constraints must distinguish mandatory conditions from preferences and support potentially conflicting requirements. Task context must preserve continuity across multiple interactions without requiring the network to reconstruct the agent's complete workflow.

Another challenge is the translation between these expressions and measurable network requirements. An agent may request that an industrial inspection be completed safely before a deadline. The infrastructure operates with measurable quantities such as latency, sensing accuracy, compute availability, data locality, reliability, and service cost. The mapping between these two levels may be uncertain, context-dependent, and dynamically updated as the task progresses.

This creates a research space around intent decomposition, capability abstraction, task representation, requirement translation, feasibility evaluation, intent negotiation, and outcome verification. A key problem is determining what information is sufficient for the network to make an effective execution decision.

### 7.3. Network Execution

Once an agent and its service relationship have been established, the next challenge is how the corresponding task is executed across networked entities and capabilities. Agentic execution may involve a single agent invoking several distributed services, or multiple agents, tools, models, network functions, and physical systems participating in different stages of the same task. These participants may cooperate, delegate, specialize, compete for resources, or operate independently while remaining associated with a common task.

The first research challenge is therefore to represent the structure of distributed agentic execution. A task may contain multiple roles, subtasks, dependencies, execution stages, and intermediate states. Some activities can proceed in parallel, while others depend on the completion or result of earlier stages. Agents may dynamically join, leave, or assume new roles during execution. ACN needs network-visible abstractions that preserve the relationships among these participants without requiring the network to reproduce the complete application workflow.

A second challenge is to bind task requirements to available capabilities. Communication, computing, sensing, data, models, tools, and external services may be provided by different entities and administrative domains. The required capabilities may also change as execution progresses. ACN must therefore determine which capabilities can support a particular execution stage, how they can be combined, and whether the resulting arrangement satisfies the task constraints. This process may involve capability discovery, service binding, task decomposition, delegation, orchestration, routing, computing placement, sensing activation, and resource allocation.

Agentic execution is also inherently dynamic. Agents may move, providers may become unavailable, network conditions may change, intermediate results may alter subsequent requirements, and policies may restrict the use of previously selected capabilities. The execution arrangement must therefore be able to evolve while preserving the identity, intent, constraints, and relevant state of the task. An important research problem is how to rebind agents, services, and infrastructure capabilities without losing task continuity.

The execution architecture itself raises an additional challenge. Some decisions require fast local responses, while others depend on task-wide information across edge, cloud, network, and service domains. A single centralized executor may face scalability, latency, and trust limitations, while fully distributed execution may make it difficult to maintain consistent task state and dependencies. Hierarchical, distributed, and hybrid execution models therefore need to be investigated, together with clear boundaries among agent decisions, network support, and infrastructure control.

Research opportunities include task and execution modeling, capability composition, multi-agent role formation, delegation, distributed orchestration, task-state synchronization, capability rebinding, task-aware routing, joint communication–computing–sensing execution, failure recovery, and task continuity. New performance metrics are also needed to characterize task completion, execution continuity, adaptation cost, recovery effectiveness, and the contribution of network support to the final task outcome.

Standardization opportunities arise at the interfaces through which distributed execution is represented and maintained. Candidate areas include task and execution identifiers, role and dependency descriptors, capability binding and invocation, task-state exchange, execution-status reporting, adaptation and rebinding triggers, cross-domain execution procedures, and provenance of actions and outcomes.

The central design principle and research challenge is:

**ACN should support distributed agentic execution while allowing agents, services, and underlying infrastructures to retain their own decision and control mechanisms.**

### 7.4. Trust, Governance, Privacy, and Accountability

Autonomous agents introduce a fundamental governance challenge because they can initiate network actions, invoke services, delegate tasks, and interact with physical systems with limited human intervention. As the degree of autonomy increases, conventional access control based on user identity and application permission alone may become insufficient.

Trust in ACN is inherently contextual. An agent that is authorized to retrieve information may not be authorized to control a machine. An agent trusted within one enterprise may require additional verification before interacting with another domain. Trust can also change during a task as authority expires, behavior deviates from expectations, or the operating environment changes.

The research problem is therefore to establish continuous and context-dependent trust without imposing prohibitive signaling and verification overhead. This includes trust establishment among previously unknown agents, delegated trust, capability-specific authorization, behavioral evidence, revocation, and cross-domain trust federation.

Privacy presents a related tension. ACN benefits from knowledge of agent identity, task association, capability requirements, and execution context, yet excessive exposure of this information can reveal sensitive user objectives, organizational operations, physical activities, or business relationships. Privacy must therefore be considered when defining every ACN information element, not added as a separate protection mechanism after the architecture is established.

Governance must also determine responsibility when autonomous actions cross multiple agents and domains. If an agent delegates a task, another agent selects a service, and the network coordinates the underlying resources, accountability cannot be established solely from packet logs. ACN needs an auditable relationship among authority, decisions, actions, and outcomes.

Research is needed on minimum disclosure, privacy-preserving capability discovery, selective context exposure, policy composition, trust federation, delegation chains, provenance, explainable coordination decisions, accountable execution and security. Standardization opportunities include trust and policy descriptors, authority credentials, delegation and revocation procedures, audit records, provenance information, and common mechanisms for expressing which agentic information may be exposed to which network functions.

A guiding principle for ACN should be:

**Agents may reason, propose, negotiate, and coordinate, but execution remains bounded by policy, authorization, feasibility, and enforcement.**

# 8. Conclusion

The emergence of autonomous agents is changing the role they play in networked systems. Agents can act on behalf of human users, organizations, machines, and infrastructure operators; express intent and invoke capabilities; interact with other agents and services; delegate and execute tasks; and maintain task state across changing execution environments. As these activities expand across devices, networks, edge and cloud platforms, and administrative domains, endpoint- and traffic-centric abstractions alone become insufficient to describe the resulting network relationships and requirements.

This white paper introduces Agentic Communication Networks (ACN) as a network-centric paradigm for the stage in which agents become first-order network entities. ACN brings three fundamental aspects of agentic operation into the scope of networking. The Agent Entity Plane provides identifiable and governable representations of agents, including their identity, principal, authority, lifecycle, and accountability. The Agentic Service Plane represents what an agent seeks to accomplish through intent, capability, context, policy, constraints, and task state. The Agentic Execution Plane supports how these service relationships are carried out across multiple agents, services, tools, network functions, and distributed communication, computing, sensing, data, and physical capabilities.

These three planes establish a clear relationship between agent intelligence and network support. Agents retain their own reasoning, planning, and application logic, while ACN provides network-facing abstractions for identity, service expression, capability discovery, task execution, trust, and policy enforcement. Such abstractions allow independently developed digital and physical agents to interact across heterogeneous platforms and infrastructures while preserving explicit authority and task context.

Looking ahead, ACN envisions a networked world in which heterogeneous digital and physical agents can discover one another, establish trusted service relationships, and execute distributed tasks across communication, computing, sensing, data, and physical infrastructures. As autonomous agents continue to multiply in number, capability, and autonomy, the network will increasingly become the common foundation that connects their identities, services, and execution processes across platforms and domains. By establishing interoperable and governable abstractions for this emerging form of communication, ACN has the potential to enable agentic systems to operate at network scale and to open a new stage in the evolution of communication networks.